\documentclass[journal]{IEEEtran}

\usepackage{lineno}
\modulolinenumbers[5]

\usepackage[colorlinks]{hyperref}
\usepackage{microtype}
\usepackage{cite}

\usepackage{tabularx}
\newcolumntype{R}{>{\raggedleft\arraybackslash}X}%
\usepackage{booktabs,multirow}
\usepackage{graphicx}
\usepackage{sidecap}

\usepackage{amsmath}
\usepackage{amssymb}
\usepackage{mathtools}
\usepackage{units}

\renewcommand{\vec}[1]{\mathbf{#1}}
\newcommand{\mat}[1]{\mathrm{#1}}

\begin{document}

\title{Learning compact $q$-space representations for multi-shell diffusion-weighted MRI}

\author{Daan Christiaens, Lucilio~Cordero-Grande, Jana~Hutter, Anthony~N.~Price, Maria~Deprez,\\ Joseph~V.~Hajnal, J-Donald~Tournier% <-this % stops a space
\thanks{Manuscript received 2~May 2018; revised 30~July 2018; accepted 24~September 2018. Please see \href{https://doi.org/10.1109/TMI.2018.2873736}{10.1109/TMI.2018.2873736} for citation information.}
\thanks{All authors are with the Centre for the Developing Brain, School of Biomedical Engineering \& Imaging Sciences, King's College London, King's Health Partners, St.\,Thomas' Hospital, London SE1\,7EH, United Kingdom (e-mail: \href{mailto:daan.christiaens@kcl.ac.uk}{daan.christiaens@kcl.ac.uk}).}% <-this % stops a space
\thanks{This research was funded by the ERC developing Human Connectome Project [Grant Agreement no. 319456] and MRC strategic funds [MR/K006355/1]. This work was additionally supported by the Wellcome/EPSRC Centre for Medical Engineering at King's College London [WT 203148/Z/16/Z] and by the National Institute for Health Research (NIHR) Biomedical Research Centre at Guy's and St Thomas' NHS Foundation Trust and King's College London. The views expressed are those of the authors and not necessarily those of the NHS, the NIHR or the Department of Health.}%
}

% Headers
\markboth{IEEE Transactions on Medical Imaging}%
{Learning compact $q$-space representations for multi-shell dMRI}

%\IEEEspecialpapernotice{(Manuscript Submitted for Review)}

%\tnotetext[abbrev]{List of Abbreviations: 
%dMRI, diffusion-weighted magnetic resonance imaging; 
%ODF, orientation distribution function; 
%RMSE, root-mean-square error; 
%SH, spherical harmonics; 
%SHARD, spherical harmonics and radial decomposition; 
%SHORE, simple harmonic oscillator based reconstruction and estimation;
%SVD, singular value decomposition.
%}

\maketitle

\begin{abstract}
Diffusion-weighted MRI measures the direction and scale of the local diffusion process in every voxel through its spectrum in $q$-space, typically acquired in one or more shells. Recent developments in microstructure imaging and multi-tissue decomposition have sparked renewed attention in the radial $b$-value dependence of the signal. Applications in motion correction and outlier rejection therefore require a compact linear signal representation that extends over the radial as well as angular domain. Here, we introduce SHARD, a data-driven representation of the $q$-space signal based on spherical harmonics and a radial decomposition into orthonormal components. This representation provides a complete, orthogonal signal basis, tailored to the spherical geometry of $q$-space and calibrated to the data at hand. We demonstrate that the rank-reduced decomposition outperforms model-based alternatives in human brain data, whilst faithfully capturing the micro- and meso-structural information in the signal. Furthermore, we validate the potential of joint radial-spherical as compared to single-shell representations. As such, SHARD is optimally suited for applications that require low-rank signal predictions, such as motion correction and outlier rejection. Finally, we illustrate its application for the latter using outlier robust regression.
\end{abstract}


%\linenumbers

%%%%%%%%%%%%%%%%%%%%%%%%%%%%%%%%%%%%%%%%%%%%%%%%%%%%%%%
%%%%%%%%%%%%%%%%%%%%%%%%%%%%%%%%%%%%%%%%%%%%%%%%%%%%%%%

\section{Introduction}

Diffusion-weighted magnetic resonance imaging (dMRI) is a noninvasive imaging technique that can probe properties of the cellular microstructure in tissue, such as neurite fibre directions in brain white matter. Its underlying principle is based on sensitizing the MRI signal to the Brownian motion of proton spins along a certain direction and scale encoded by the diffusion gradient \cite{LeBihan1986}. When sampled across a dense set of narrow gradient pulses, the resulting dMRI signal measures the 3-D Fourier spectrum of the ensemble average propagator of the diffusion process in each voxel. This spectral domain is known as $q$-space, in analogy with the $k$-space concept in conventional MRI \cite{Callaghan1988}.

Most dMRI data are nowadays acquired with dense sampling over one or more shells in $q$-space, with radius determined by the $b$-value of each shell (assuming fixed diffusion time) \cite{LeBihan1986}. Such single- and multi-shell protocols facilitate a straightforward trade-off between high angular and radial resolution; the former being advantageous for tractography, the latter for microstructure imaging. Moreover, the signal in each shell can be efficiently represented in the spherical harmonic (SH) basis \cite{Frank2002}, which linearizes spherical deconvolution and related techniques \cite{Tuch2002, Tuch2004, Anderson2005, Tournier2004, Tournier2007, Descoteaux2009}. Recent developments in microstructure modelling \cite{Assaf2005, Assaf2008, Alexander2010, Fieremans2011, Zhang2012, Jespersen2012, Reisert2014, Daducci2015, Kaden2016, Reisert2017, Novikov2018a} and multi-tissue decomposition \cite{Jeurissen2014, Christiaens2015a, Christiaens2017} have sparked renewed attention in the radial $b$-value dependence of the signal. There is therefore a need for compact, linear representations of the dMRI signal that extend over the radial as well as angular domain.

Applications such as motion and distortion correction \cite{Rohde2004, Andersson2016} and outlier rejection \cite{Chang2005, Pannek2012, Tobisch2016} require such signal representations for generating rank-reduced predictions to which the input data can be registered or compared. Variational methods for (super-resolution) image reconstruction \cite{Scherrer2012} can also benefit from linear and compact representations as these simplify the required numeric optimization. Furthermore, compact signal representations may prove useful for image denoising \cite{Veraart2016} and for multi-site data harmonization \cite{Mirzaalian2016}. Crucially, signal representations in such applications should be free of biophysical assumptions, as not to affect subsequent analysis and enable comparison of different modelling approaches.

Here, we introduce a model-free representation of the $q$-space signal based on the spherical harmonics basis within each shell and a linear, orthonormal decomposition in the radial domain. The combined \emph{spherical harmonics and radial decomposition} (SHARD) offers a bespoke signal representation across all shells, hence building a data-driven basis for the $q$-space signal in the images at hand.

This approach is similar in spirit to other blind source separation methods in dMRI, such as sparse or convex nonnegative spherical factorization \cite{Reisert2013, Christiaens2015b, Christiaens2017}. However, while nonnegativity constraints of the factorized orientation distribution functions (ODF) in those techniques are motivated on biophysical grounds, they also---by necessity---lead to increased residuals on the signal representation that may still contain relevant structure. This work does not impose any constraints apart from orthogonality, sacrificing direct biological interpretability in favour of a model-free, unconstrained basis that is better suited for low-rank signal approximations needed in motion correction and related applications, and which forms an effective basis for subsequent, more biologically inspired analyses.

In contrast to the harmonic oscillator basis \cite{Ozarslan2013, Fick2016} and other similar representations \cite{Assemlal2009, Descoteaux2011, Caruyer2012, Merlet2011, Merlet2013, Hosseinbor2013, Rathi2014}, the SHARD basis does not depend on scaling parameters and is implicitly calibrated to the data at hand. As opposed to the cumulant expansion \cite{Kiselev2010}, higher-order tensor representations \cite{Ozarslan2003, Liu2004, Schultz2014}, or Gaussian processes \cite{Andersson2015}, our decomposition is directly compatible with the SH basis and hence inherits its mathematical properties, including its ability to linearize spherical deconvolution. 

In this paper, we outline the theoretical underpinnings for the SHARD decomposition and evaluate its representation accuracy in comparison to alternative bases in human brain data with shells spanning across a range of $0 \leq b \leq \unitfrac[10]{ms}{\mu\!m^2}$. Additionally, we illustrate its use in an example practical application for outlier rejection.

%%%%%%%%%%%%%%%%%%%%%%%%%%%%%%%%%%%%%%%%%%%%%%%%%%%%%%%

\section{Theory}\label{sec:theory}

For fixed diffusion time $\tau$, the diffusion-weighted signal $S(\vec{q})$ at encoding $\vec{q} \in \mathbb{R}^3$ can be represented in a linear basis as
\begin{equation}
	S(\vec{q}) = \sum_{j=1}^\infty c_j \, \Psi_j(\vec{q}) \quad .
\end{equation}
In this work, we seek a complete, orthonormal basis $\Psi_j(\vec{q})$ suitable for unconstrained, low-rank representations of the dMRI signal. In addition, we require the basis functions to be separable in spherical coordinates, and adopt the basis of real symmetric spherical harmonics (SH) $Y_\ell^m(\theta, \phi)$ for the angular part:
\begin{equation}
	\Psi_j(\vec{q}) = R_\ell^n(q) \, Y_\ell^m(\hat{\vec{q}}) \quad ,
\end{equation}
where $q = \| \vec{q} \|$, $\hat{\vec{q}} = \vec{q}/q$, and index $j$ is determined by even SH order $\ell$, phase $m \in [-\ell,\ell]$, and radial basis degree $n$. Crucially, the basis functions $R_\ell^n(q)$ for the radial domain depend on the SH order $\ell$ but not on the phase $m$. This ensures that the radial basis can vary between independent SH frequency bands, while remaining invariant to the orientation of the dMRI signal.

Related work has presented different choices of $R_\ell^n(q)$. Hosseinbor~et~al.~\cite{Hosseinbor2013} used the spherical Bessel functions $j_\ell(\alpha\,q)$ at frequency $\alpha$, which arise naturally when casting the $q$-space inverse Fourier transform in spherical coordinates. However, discretizing the frequency $\alpha$ requires imposing zero-value or zero-derivative Sturm-Liouville boundary conditions at $q_\text{max}$ \cite{Wang2009}, which are generally incompatible with the nature of the dMRI signal and can thus cause aliasing effects. \"Ozarslan~et~al.~\cite{Ozarslan2013} introduced the harmonic oscillator (SHORE) basis of spherical Laguerre polynomials as a higher-order generalization of the diffusion tensor model, with basis functions depending on a scale factor $\zeta$ that needs to be tuned to the voxel or image. Finally, many current multi-shell analysis methods that treat shells independently can be regarded as using Dirac-delta basis functions $R_\ell^n(q) = \delta(q - q_n)$ at discrete shell positions $q_n$ \cite{Jeurissen2014, Christiaens2017, Reisert2017}.

Here we propose to learn an orthonormal function basis for the radial domain from the data, using the singular value decomposition (SVD) of the signal across different shells and in different SH frequency bands $\ell$. Because the SH basis functions $Y_\ell^m$ are mutually orthogonal between different orders $\ell$, it suffices to ensure orthogonality between basis functions $R_\ell^n$ of different $n$ for fixed $\ell$. Hence, we can independently resolve SVD responses in each band $\ell$. To this end, we arrange the order-$\ell$ SH coefficients of all voxels in the image, mask or patch, in a $N_\text{shells} \times (N_\text{vox} \cdot (2\ell+1))$~matrix
\begin{equation}\label{eq:matstruct1}
	\mat{S}_\ell = \begin{bmatrix}
		\mat{S}_\ell^{(1)} & \mat{S}_\ell^{(2)} & \cdots & \mat{S}_\ell^{(N_{vox})}
	\end{bmatrix}
\end{equation}
with
\begin{equation}\label{eq:matstruct2}
	\mat{S}_\ell^{(v)} = \begin{bmatrix}
		s_{\ell,q_1,v}^{-\ell} & \cdots & s_{\ell,q_1,v}^{m} & \cdots & s_{\ell,q_1,v}^{\ell} \\
		\vdots & & \vdots & & \vdots \\
		s_{\ell,q_N,v}^{-\ell} & \cdots & s_{\ell,q_N,v}^{m} & \cdots & s_{\ell,q_N,v}^{\ell}
	\end{bmatrix}	\quad,
\end{equation}
where $s_{\ell,q,v}^m = \langle Y_\ell^m(\theta, \phi) , S_v(q, \theta, \phi) \rangle$ is the projection of the signal onto the SH basis, i.e., the $\ell,m$ SH coefficient of shell $q$ in voxel $v$. Thus, shells $q$ are laid out one per row, voxels $v$ and phase terms $m$ are laid out in columns. The truncated SVD decomposes this matrix
\begin{equation}
	\mat{S}_\ell = \sum_{n=1}^{N_\text{shells}} \vec{u}_{\ell,n} \, \sigma_{\ell,n} \, \vec{v}_{\ell,n}^\top = \mat{U}_\ell \, \Sigma_\ell \, \mat{V}^\top_\ell
\end{equation}
into left and right singular vectors $\vec{u}_{\ell,n}$ and $\vec{v}_{\ell,n}$, and associated singular values $\sigma_{\ell,n}$ in decreasing order. The vectors $\vec{u}_{\ell,n}$ (eigenvectors of $\mat{S}_\ell \, \mat{S}_\ell^\top$) span an orthonormal basis for the order-$\ell$ SH band across shells, that uniquely determine the rotation-invariant principal components in the signal. The vectors $\vec{v}_{\ell,n}$ contain the associated voxel coefficients of each basis component. The singular values $\sigma_{\ell,n}$ measure the effect size of each component.

\begin{figure*}
	\centering
	\includegraphics[width=.9\textwidth]{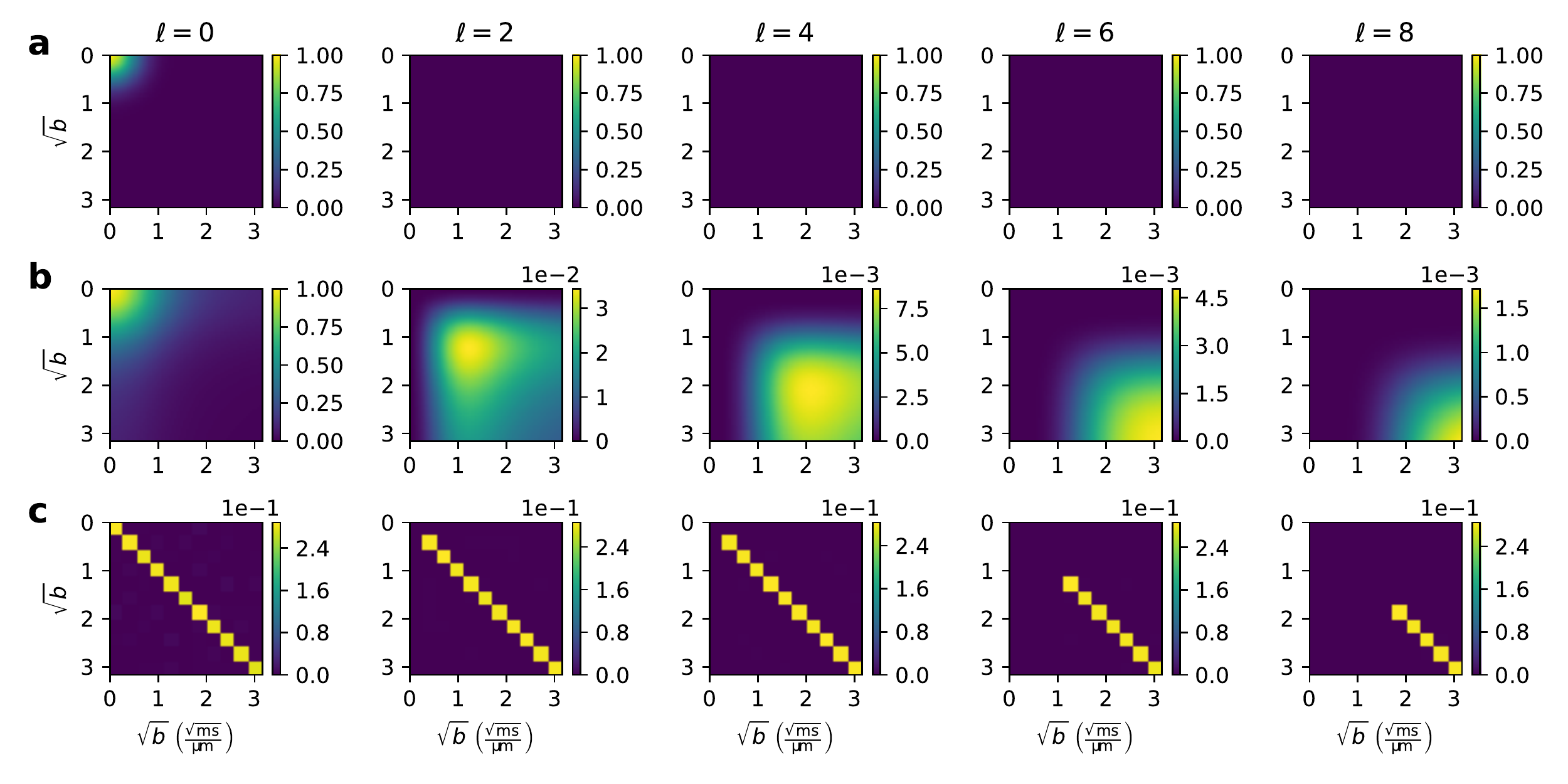}
	\caption{Covariance matrices $\tilde{\mat{S}}_\ell \, \tilde{\mat{S}}_\ell^\top$ in simulated signals. (a) Simulated free water signal with isotropic diffusivity $D_f = \unitfrac[3.0]{\mu\!m^2}{ms}$. (b) Simulated bi-exponential signal with 50\% restricted intra-axonal diffusion with $D^{\parallel}_i = \unitfrac[2.2]{\mu\!m^2}{ms}$ and 50\% hindered extra-axonal diffusion with $D^{\parallel}_e = \unitfrac[1.2]{\mu\!m^2}{ms}$ and $D^{\bot}_e = \unitfrac[0.7]{\mu\!m^2}{ms}$. (c) Simulated effect of Gaussian noise on the $q$-space sampling scheme of dataset~1 after rescaling with the number of samples, reducing to the identity matrix in all overdetermined shells.}
	\label{fig:simulation}
\end{figure*}

This decomposition directly extends the isotropic \mbox{($\ell = 0$)} data representation used in Tournier~et~al.~\cite{Tournier2015} for optimizing the diffusion gradient sampling scheme to higher SH order $\ell \geq 0$, thus facilitating a generalized signal representation including all angular information. The proposed basis also extends the rotation-invariant signal features that have recently been introduced for microstructural modelling \cite{Kaden2016, Novikov2018a, Reisert2017}. In a single voxel, the widely accepted spherical convolution model \cite{Tournier2004, Jespersen2007, Novikov2018a} assumes that the signal can be factorized into an axially symmetric response function $h_{\ell,q}$ (the microstructure) and an orientation distribution function $p_\ell^m$ (ODF, the mesostructure): $s_{\ell,q}^m = h_{\ell,q}\,p_\ell^m$. Under this identity, the voxel-wise covariance matrix
\begin{align}
	\mat{S}_\ell^{(v)} {\mat{S}_\ell^{(v)}}^\top &= \begin{bmatrix}
		\sum_m {s_{\ell,q_1}^m}^2  & \cdots & \sum_m s_{\ell,q_1}^m s_{\ell,q_N}^m \\
		\vdots & \ddots & \vdots  \\
		\sum_m s_{\ell,q_N}^m s_{\ell,q_1}^m & \cdots & \sum_m {s_{\ell,q_N}^m}^2 
	\end{bmatrix} 
	\label{eq:voxcovar} \\
	&= \begin{bmatrix}
		h_{\ell,q_1}^2  & \cdots & h_{\ell,q_1} h_{\ell,q_N} \\
		\vdots & \ddots & \vdots  \\
		h_{\ell,q_N} h_{\ell,q_1} & \cdots & h_{\ell,q_N}^2 
	\end{bmatrix}  \underbrace{\left(\textstyle\sum_m {p_\ell^m}^2\right)}_{p_\ell}
	\label{eq:voxevd} \\
	&= p_\ell \, \vec{h}_\ell \vec{h}_\ell^\top \quad \text{with} \quad \vec{h}_\ell = [h_{\ell,q_1} \cdots h_{\ell,q_N}]^\top
\end{align}
becomes a rank-1 matrix (in the absence of noise) scaled with the ODF power $p_\ell$. The diagonal elements of this matrix \cite{Kazhdan2003} or ratios thereof \cite{Reisert2017} have previously been used as rotation-invariant signal features. Here, we use the full covariance matrix $\mat{S}_\ell^{(v)} {\mat{S}_\ell^{(v)}}^\top$, whose principal eigenvector $\vec{h}_\ell / \|\vec{h}_\ell\|$ captures the microstructural information in the signal up to a scaling factor. Example covariance matrices for (isotropic) free water and bi-exponential white matter are shown in Fig.~\ref{fig:simulation}a-b, simulated directly in spherical harmonics using the forward relation for Gaussian signal response in each compartment~\cite{Anderson2005}.

\begin{figure*}[t]
	\centering
		\includegraphics[width=\textwidth]{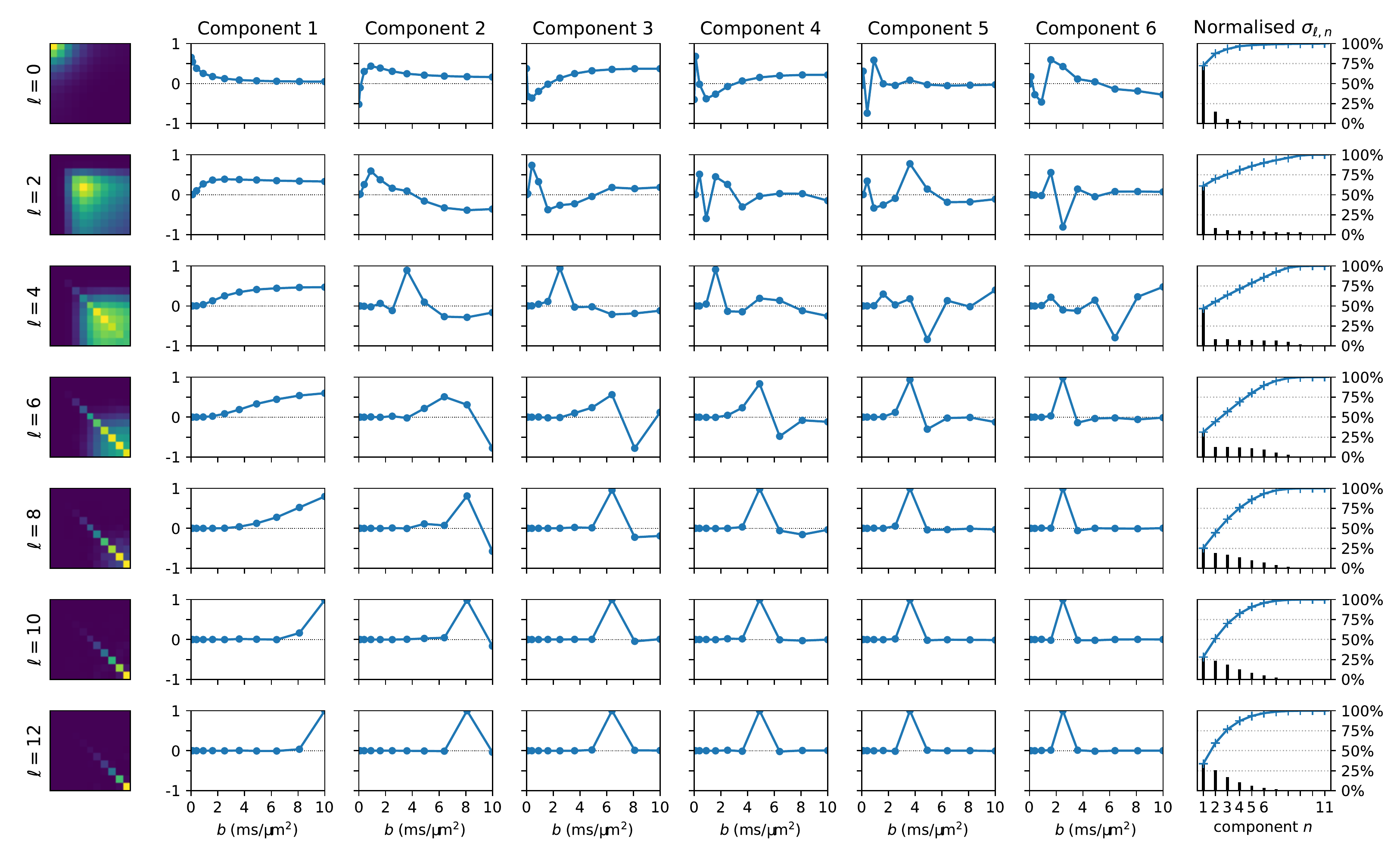}
	\caption{SHARD decomposition of dataset~1. Rows correspond to spherical harmonic bands of order $\ell = 0, 2, \ldots, 12$ The left column illustrates square, symmetric matrices $\tilde{\mat{S}}_\ell \, \tilde{\mat{S}}_\ell^\top$, showing the covariance between shells in each harmonic band. The middle columns plot the decomposition basis vectors of the 6 leading components, corresponding to the principal eigenvectors of the matrices on the left. The plots in the right column depict the singular values of each component in decreasing order and their cumulative sum.}
	\label{fig:basis}
\end{figure*}

When extended across all voxels, the full covariance matrix $\mat{S}_\ell\,\mat{S}_\ell^\top = \sum_v \mat{S}_\ell^{(v)} {\mat{S}_\ell^{(v)}}^\top$ is no longer rank-1 and its eigenvectors capture the spatial variation of $h_{\ell,q}$. Therefore, the basis $\mathcal{U} = \{\mat{U}_\ell \,|\, \forall\, \ell \geq 0\}$ for the radial domain spans the complete microstructural properties encoded in the signal for a given multi-shell protocol. Example matrices $\mat{S}_\ell\,\mat{S}_\ell^\top$ and derived basis functions and effect sizes are illustrated in Fig.~\ref{fig:basis}.

In the remainder of this work, we define basis functions $R_\ell^n(q)$ for the radial domain as discrete functions at shell locations $q_i$, i.e.,
\begin{equation}\label{eq:interp}
	R_\ell^n(q) = \vec{u}_{\ell,n} \cdot \left[\,\delta(q - q_1)\,,\,\ldots\,,\, \delta(q - q_N)\,\right]^\top \quad.
\end{equation}
The functions of smallest degree $n \leq N_\text{shells}$ capture the strongest covariance between individual shells across all voxels in the image. Note that in general, one may define $R_\ell^n(q)$ using any desired numerical interpolation method between shells.

By virtue of the Eckart-Young SVD theorem, this basis provides the optimal low-rank representation of multi-shell dMRI data: a signal representation in a truncated basis of rank $r$ has minimum Frobenius error w.r.t. the full-rank data. To this end, we can define index $j(\ell,m,n)$ in such a way that all basis functions are sorted in order of decreasing effect size $\varepsilon_j = \frac{\sigma_{\ell,n}^2}{2\ell+1}$, where the normalisation factor accounts for the number of basis functions in each harmonic band $\ell$.
As such, the rank-$r$ approximation
\begin{equation}
	\hat{S}(\vec{q}) = \sum_{j=1}^r c_j \, \Psi_j(\vec{q})
\end{equation}
represents the multi-shell dMRI signal with minimal $L_2$-error $\| S(\vec{q}) - \hat{S}(\vec{q}) \|_{L_2}$. This property facilitates applications for denoising, motion correction and outlier rejection.

%%%%%%%%%%%%%%%%%%%%%%%%%%%%%%%%%%%%%%%%%%%%%%%%%%%%%%%

\section{Materials and Methods}

\subsection{Data and preprocessing}

\subsubsection{Dataset 1} Multi-shell data were acquired in a healthy subject on a \unit[3]{T} Philips Achieva TX system using a 32-channel head coil and image-based shimming. The dMRI data were sampled over 11 shells ranging from $b = \unitfrac[0]{s}{mm^2}$ to $b = \unitfrac[10,\!000]{s}{mm^2}$, spaced equidistantly in $q$ (note that we use $q \propto \sqrt{b}$ interchangeably throughout this paper). The number of samples in each shell ranges from 15 to 115 respectively, scaled linearly with the shell surface $\propto q^2 \propto b$, resulting in a total of 550 dMRI volumes interleaved for optimal duty cycle effects \cite{Hutter2017}. Each volume was acquired with single-shot echo planar imaging (EPI), $G_\text{max} = \unitfrac[80]{mT}{m},$ TE = \unit[99]{ms}, TR = \unit[6900]{ms}, multiband factor MB = 2, SENSE factor 2, isotropic resolution \unit[2.5]{mm}. In addition, $b=0$ single-band reference scans were acquired with 4 phase encoding directions, for image reconstruction \cite{CorderoGrande2016ismrm} and for susceptibility-induced distortion correction.

\subsubsection{Dataset 2} A second healthy volunteer was scanned on the same system with higher spatial resolution and more standard $b$-value range. EPI acquisition parameters are: MB = 3, SENSE = 2, isotropic resolution \unit[2.0]{mm}, $\text{TE} = \unit[105]{ms}$ and $\text{TR} = \unit[6000]{ms}$. The dMRI gradient table was designed with shells at $b = 0$, 600, 1400, 2600 and $\unitfrac[4000]{s}{mm^2}$, with 8, 12, 28, 52 and 80 samples per shell respectively (180 gradient directions in total), in this case with AP/PA phase encoding.

\subsubsection{Preprocessing} All data were preprocessed with a patch-based image denoising technique based on the Marchenko-Pastur distribution in the complex (phase-magnitude) image domain coupled with phase-corrected real reconstruction \cite{Veraart2016, Gavish2017}, Gibbs-ringing removal \cite{Kellner2016}, and motion and distortion correction \cite{Andersson2003, Andersson2016}. Dataset~1 was also corrected for signal drift due to gradient heating over the longer acquisition time \cite{Vos2016}. Brain masking was done on the mean $b=0$ image \cite{Smith2002}.

\subsection{SHARD basis construction}

The SHARD basis functions for multi-shell dMRI are learned from the per-shell signal coefficients in spherical harmonics. The first step in the basis construction is therefore to project each acquired shell onto the real, even-order SH basis. When the number of acquired samples varies between shells, as is the case in our data, a direct overdetermined SH fit per shell can force $\ell_\text{max}$ to differ between shells. Although such direct fit is sufficient and entirely compatible with the theory of Section~\ref{sec:theory}, in this work we prefer to incorporate the Laplace-Beltrami regularizer $\frac{1}{q^2} \Delta_{\mathcal{S}^2} \propto -\frac{1}{b} \ell (\ell+1)$ in order to enable working at fixed $\ell_\text{max}$ across all shells, including underdetermined cases. In this way, the basis functions will span all shells for all orders $\ell$. The SH coefficients of each shell are then obtained with regularized least squares
\begin{equation}\label{eq:shfit}
	\vec{s}_b^\ast = \arg \min_\vec{s} \, \tfrac{1}{n_b} \, \| \vec{y}_b - \mat{Y}_b \, \vec{s} \|^2 + \gamma^2 \, \| \mat{L}_b \, \vec{s} \|^2 \quad ,
\end{equation}
where $\vec{y}_b$, $\vec{s}_b$, and $n_b$ are respectively the signal intensities, the SH coefficients, and the number of samples in shell $b$. $\mat{Y}_b$ is the corresponding SH basis matrix, and $\mat{L}_b$ is a diagonal matrix that contains the Laplace-Beltrami regularization factors $-\frac{1}{b} \ell (\ell+1)$. The regularizer weight $\gamma$ is tuned empirically and kept small enough not to distort the power spectrum of an unregularized fit on the outer (overdetermined) shell (Fig.~\ref{fig:shpower}), yielding values $\gamma = \unitfrac[0.02]{\mu\!m^2}{ms}$ for dataset~1 and $\gamma = \unitfrac[0.005]{\mu\!m^2}{ms}$ for dataset~2. We underscore that regularization is only used in the basis construction, not in the subsequent multi-shell fit. The SH order $\ell_\text{max} = 12$ in dataset~1 and $\ell_\text{max} = 8$ in dataset~2, were set to the maximum for which at least two shells are overdetermined. 

The resulting SH coefficients of all shells (including $b=0$) and all voxels within a brain mask are subsequently arranged in matrices $\mat{S}_\ell$ with structure given by equations~\eqref{eq:matstruct1}--\eqref{eq:matstruct2}. To normalise the noise variance with respect to the number of samples in each shell, the signal is rescaled with the square root of the number of samples on each shell. To this end, a diagonal weighting matrix $\mat{W} = diag(\cdots \sqrt{n_b} \cdots)$ is introduced. The SVD of the reweighted matrices $\tilde{\mat{S}}_\ell = \mat{W}\,\mat{S}_\ell = \tilde{\mat{U}}_\ell \, \Sigma_\ell \, \mat{V}_\ell$ provides the basis matrices $\mat{U}_\ell = \mat{W}^{-1}\,\tilde{\mat{U}}_\ell$ for radial $q$-space. These correspond to the eigenvectors of the weighted covariance matrices $\mat{W}\,\mat{S}_\ell\,\mat{S}_\ell^\top\,\mat{W}^\top$. As shown in Fig.~\ref{fig:simulation}c, this normalisation ensures uniform variance across all shells under independent simulated noise, regardless of the number of samples per shell.

We explore two possible strategies for selecting rank-$r$ subsets of the complete basis $\mathcal{U}$, based on the observation that the strongest signal contribution originates from low-$\ell$ and low-$n$ components. In the first strategy, the number of basis functions per harmonic order is matched to the Bessel and SHORE bases (next section) by selecting components in upper-left triangles in Fig.~\ref{fig:basis}, for example using 3 basis vectors at $\ell=0$, 2 at $\ell=2$ and 1 at $\ell=4$. Since each component in band $\ell$ contributes $2\ell+1$ SH phase terms, in this example the total basis rank $r=3 \times 1 + 2 \times 5 + 1 \times 9=22$. We will refer to this strategy as \emph{matched} ordering. In the second strategy, the basis functions are optimally ordered for decreasing effect size $\varepsilon_j$ as described in Sec.~\ref{sec:theory} and illustrated in Suppl.\,Tables~\ref{tbl:effect1}--\ref{tbl:effect2}. We will refer to this strategy as \emph{optimal} ordering.

\subsection{Comparison with Bessel and SHORE bases}

As explained in Sec.~\ref{sec:theory}, SHARD learns the orthonormal rank-$r$ basis that best represents the $q$-space signal in the data at hand. Here, we wish to verify if SHARD indeed outperforms alternative bases and by what margin. To this end, we evaluate and compare the residual root-mean-squared error (RMSE) in a rank-$r$ fit with equal rank representations in the spherical Bessel function basis and the SHORE basis.

The spherical Bessel functions $j_\ell(\alpha\,q)$ provide a basis for the Fourier transform in spherical coordinates. Here, we discretize the radial frequency $\alpha$ using a zero-derivative boundary condition at $q_\text{max}$:
\begin{equation}
	R_\ell^n(q) = \frac{1}{N_\ell^n} \, j_\ell(\alpha_\ell^n\,q) \quad,
\end{equation}
where $\alpha_\ell^n = x_\ell^n / q_\text{max}$ with $x_\ell^n$ the positive zeros of $j'_\ell(x)$ and $N_\ell^n$ is a normalization factor. Basis functions are ordered for increasing energy $\propto \alpha_\ell^n$ \cite{Wang2009}. For the low basis rank used in our experiments, this minimum-energy ordering comes down to selecting $2(n-1)+\ell \leq \ell_\text{max}$.

The SHORE basis is defined using the associated Laguerre polynomials $L_n^p(x)$:
\begin{equation}
	R_\ell^n(q) = \frac{1}{N_\ell^n} \, \left(\frac{q^2}{\zeta}\right)^{\ell/2} \exp\!\left(\frac{-q^2}{2\zeta}\right) \, L_{n-\ell}^{\ell+1/2}\!\left(\frac{q^2}{\zeta}\right) \quad,
\end{equation}
where $N_\ell^n$ is a normalization factor \cite{Merlet2013}. Here, we calibrate the scale parameter $\zeta$ to each dataset using non-linear minimization of the RMSE across the full brain mask, in order to ensure the best possible data representation. For real, symmetric signals, the index $n$ ranges from $\ell/2$ to $\ell_\text{max}/2$ \cite{Ozarslan2013}.

\subsection{Rank selection}

SHARD defines the effect size of each basis component, but does not determine what effect size is relevant in a given application. A higher basis rank will yield a more accurate representation of the data, at the expense of having more parameters to fit. Selecting the optimal trade-off will inevitably depend on the application. Here, we primarily consider applications that require dMRI contrast prediction, such as outlier rejection and motion correction. In this case, one possible strategy for rank selection is leave-one-out cross-validation, in which we verify the capacity of a basis of certain rank $r$ to accurately predict unseen data samples. We then select the basis rank for which the prediction error is minimal.

\subsection{Example application in outlier rejection}

We illustrate a potential application of the SHARD basis for outlier robust regression from M-estimator theory \cite{Huber1981}, using iteratively reweighted least squares fitting with Soft-$L_1$ and Cauchy loss functions \cite{Holland1977}. This scheme starts with an initial least-squares estimate in each voxel, and iteratively downweights samples with large residuals to reduce the effect of outliers. The sample weights are calculated as $w(e) = \rho'(e)/e$, according to a chosen loss function
\begin{equation}
  \rho(e) = \begin{dcases}
  			e^2  			& \text{$L_2$-loss} \\
                      	2\sqrt{1+e^2} - 2 & \text{Soft-$L_1$ loss} \\
			\log(1+e^2) 	& \text{Cauchy loss}
		\end{dcases} \quad,
\end{equation}
where $e$ is the sample residual, rescaled by a normalisation factor set to the median absolute deviation in the sample.

We evaluate the performance of each loss function in low-rank multi-shell representations in data samples with varying amount of simulated outliers. A random subsample of 1000 voxels was extracted from the data in the brain mask. For any percentage of outliers, ranging from 0\% to 25\%, outliers are simulated by setting a randomly chosen subset of the samples in each voxel to zero. As such, we mimic at the individual voxel level the effect of slice dropouts, arguably the predominant source of outliers in EPI. We then evaluate the RMSE between the uncorrupted data samples and their predictions in linear and robust least-squares regression.

%%%%%%%%%%%%%%%%%%%%%%%%%%%%%%%%%%%%%%%%%%%%%%%%%%%%%%%

\begin{figure*}[t]
	\centering
	\raisebox{3.5cm}{\rotatebox[origin=c]{90}{\sffamily\bfseries(a) Dataset 1}}
	\includegraphics[width=.9\textwidth]{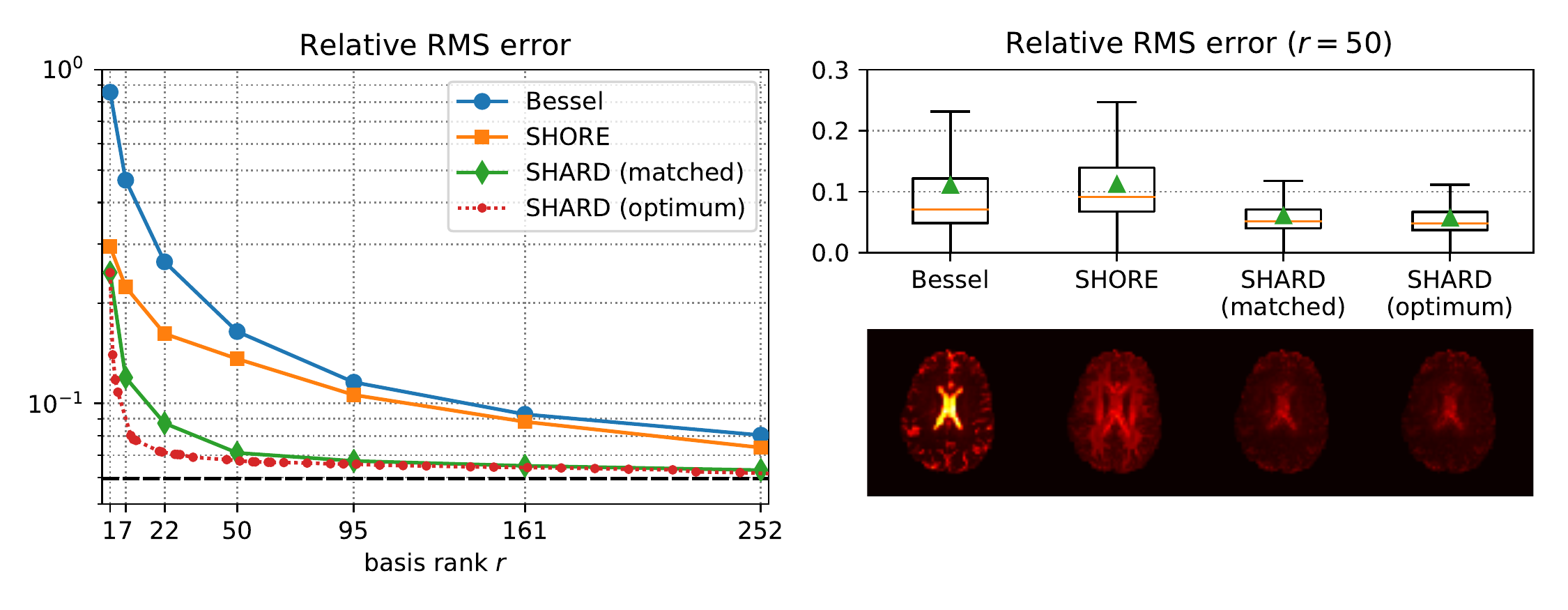} \\
	\raisebox{3.5cm}{\rotatebox[origin=c]{90}{\sffamily\bfseries(b) Dataset 2}}
	\includegraphics[width=.9\textwidth]{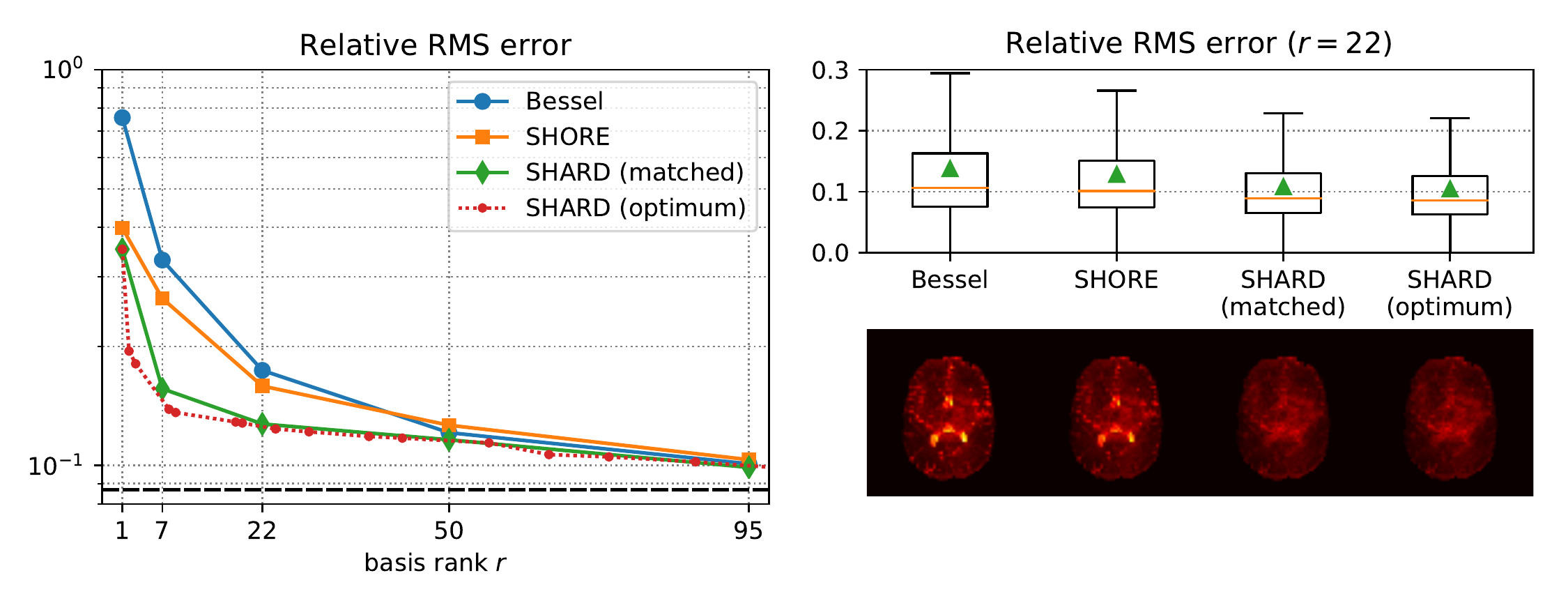}
	\caption{Root mean squared error (RMSE) of the signal fit in the different bases, shown for dataset~1 (a) and~2 (b). The graphs on the left depict the total RMSE across the brain mask, relative to the signal power, on a logarithmic scale. This fitting error reduces for increasing basis rank $r$. For SHARD, the RMSE converges to the residual fitting error of an independent SH fit, indicated by the black dashed line in the graph. On the right, box plots of the voxel-wise relative RMSE are shown at rank $r = 50$ for dataset~1 and at rank $r=22$ for dataset~2, as well as images of its value across the brain. The green triangles indicate the mean RMSE.}
	\label{fig:rmse}
\end{figure*}

\section{Results}

\subsection{SHARD basis construction and evaluation}

The SHARD basis functions are plotted in Fig.~\ref{fig:basis} for dataset~1 and in Suppl.\,Fig.~\ref{fig:basis2} for dataset~2. These basis functions capture the covariance of the data between shells. Within each harmonic band $\ell$, subsequent components capture higher radial frequencies of decreasing effect size. Across harmonic bands, the covariance between shells reduces with increasing $\ell$. At $\ell \geq 10$, this results in near-independent shells: $\tilde{\mat{S}}_\ell \, \tilde{\mat{S}}_\ell^\top$ reduces to a near-diagonal matrix and basis functions for $n \geq 2$ reduce to Dirac \mbox{$\delta$-functions}. The effect sizes of all components (Suppl.\,Tables~\ref{tbl:effect1}-\ref{tbl:effect2}) are largest for the basis functions of low $\ell$ and $n$, plotted in the upper-left triangle in Fig.~\ref{fig:basis}. A similar observation can be made in the magnitude images of all components (Suppl.\,Fig.~\ref{fig:svdcomponents} and~\ref{fig:svdcomponents2}).

We evaluate the RMSE of the signal fit in the matched and optimal SHARD bases, normalised to the RMS power of the measured signal, and compare this measure to other bases. Figure~\ref{fig:rmse} plots the relative RMSE as a function of the basis rank $r$. As expected, the RMSE decreases with increasing $r$. We observe that SHARD outperforms the calibrated SHORE and Bessel bases at any rank, and most strongly in the low-rank range ($r < 50$). In dataset~1, we also found reduced residuals at high basis rank, most likely due to stronger non-monoexponential signal decay in the extended $b$-value range, which is more efficiently captured by SHARD. Towards high rank, the RMSE in SHARD converges to the residual fitting error of the independent SH fit from which the basis was derived. This residual SH fitting error is indicated by the black dashed line and sets a lower bound on the SHARD RMSE. We also see that the ordering strategy of the SHARD basis functions has a minor effect on accuracy in the low-rank regime, with selection according to decreasing component effect size slightly outperforming the matched order. The images on the right in Fig.~\ref{fig:rmse} show that highest fitting errors occur in the ventricles due to physiological noise, and also in regions affected by an unfolding artefact in dataset~2. SHARD provides more accurate signal representation with lower and more homogenous fitting errors across the brain.

\subsection{Rank selection}

Leave-one-out cross-validation in the optimally-ordered SHARD basis selected rank $r=46$ in dataset~1 and $r=19$ in dataset~2 (Suppl.\,Tables~\ref{tbl:effect1}--\ref{tbl:effect2}). When the component order is matched to the SHORE and Bessel bases, these values are rounded to $r=50$ for dataset~1 and $r=22$ for dataset~2. The best dMRI contrast prediction is thus obtained at fairly low rank, where SHARD is shown to outperform alternatives.

\begin{figure}[t]
	\centering
	\includegraphics[width=.9\columnwidth]{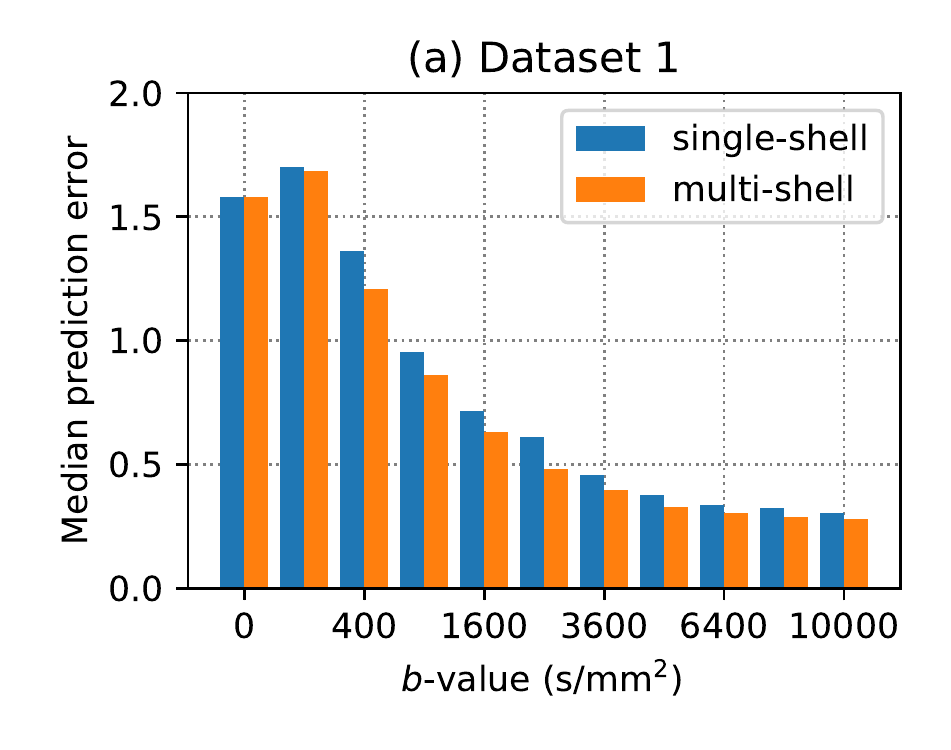}\\
	\includegraphics[width=.9\columnwidth]{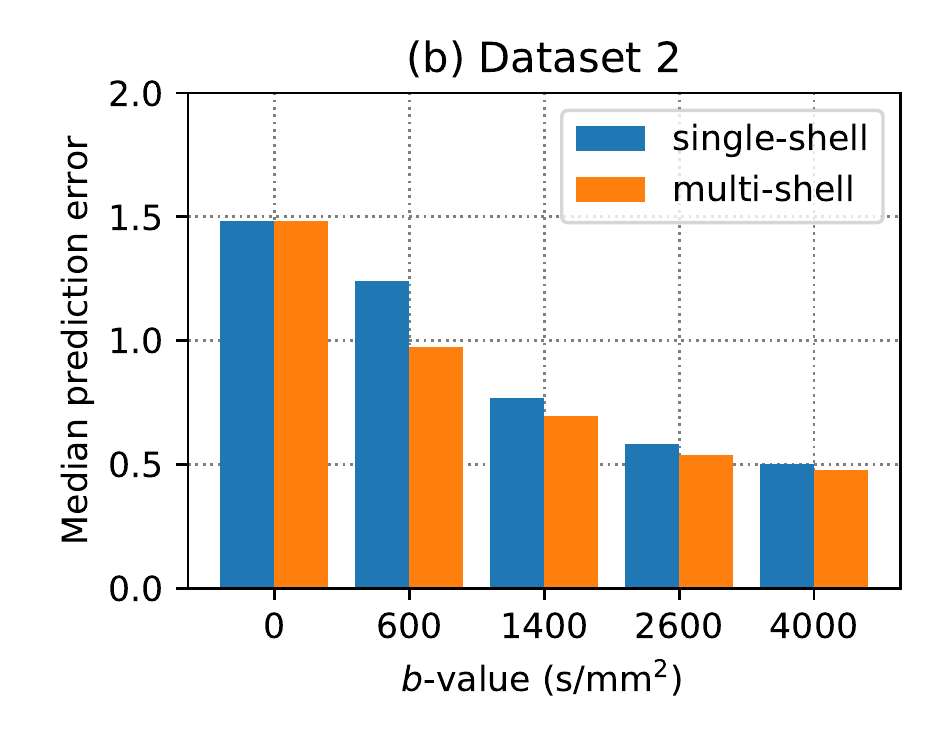}
	\caption{Comparison of the single-shell and multi-shell prediction error in nested leave-one-out cross-validation for datasets~1~(a) and~2~(b). The blue bars show the median prediction error for single-shell SH cross-validation, where each dMRI volume is predicted from all other volumes in the shell. The orange bars show the median prediction error for multi-shell SHARD cross-validation, where each dMRI volume is predicted from all other volumes across all shells.}
	\label{fig:pred}
\end{figure}

In addition, we compare SHARD to the per-shell SH basis by exploring the prediction error in nested leave-one-out cross-validation of single-shell and multi-shell predictions. In the SH basis, we generate a prediction for every dMRI volume from all other volumes in the same shell. In the SHARD basis, the prediction for each volume is generated from all other dMRI volumes across all shells. In both cases, the basis rank is selected to yield lowest overall prediction error per shell using cross-validation on the remaining dMRI volumes. As such, this experiment compares the best achievable prediction error in a single-shell setup, to the minimal prediction error in a multi-shell setup. The plots in Fig.~\ref{fig:pred} demonstrate that a multi-shell prediction consistently outperforms a single-shell approach, with improvements ranging from 4\% to 20\%. This shows the potential interest of using multi-shell signal representations in motion and distortion correction applications.

\begin{figure*}[t]
	\centering
	\includegraphics[width=\textwidth]{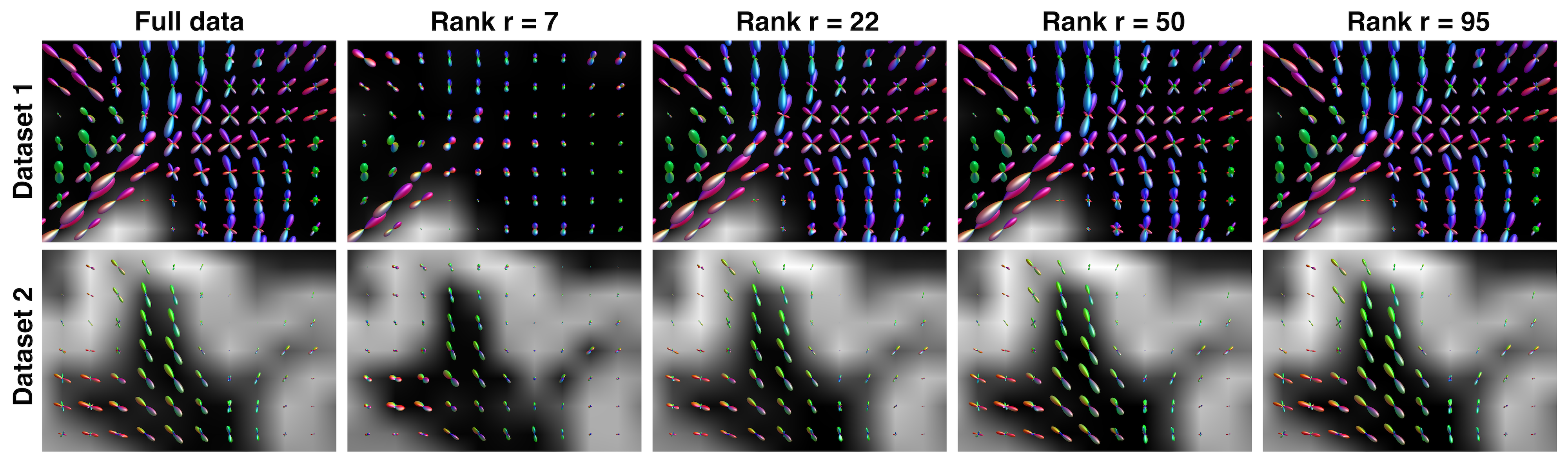}
	\caption{Fibre orientation distribution functions in the full data (left column), and rank-reduced data in the matched SHARD basis. The top row depicts the centrum semiovale in dataset~1, overlaid onto the CSF tissue fraction. The bottom row shows the superior frontal gyrus in dataset~2, overlaid onto the grey matter fraction.}
	\label{fig:odfs}
\end{figure*}

The low-rank projected data inevitably sacrifices part of the information in the input. We qualitatively investigate this information loss in rank-reduced data using unsupervised multi-tissue spherical factorization under convexity and nonnegativity constraints into 3 components of SH order (8, 0, 0), respectively associated with white matter, grey matter, and cerebrospinal fluid \cite{Christiaens2017}. The resulting fibre ODFs are depicted in Fig.~\ref{fig:odfs} for the full data and for rank-$r$ projected data. We observe that the data quality indeed improves for increasing rank. A minimum rank $r=22$ (corresponding with SH order $\ell_\text{max}=4$) is required for a faithful data representation in crossing fibre regions. From rank $r \geq 50$, the ODF factorization becomes nearly indistinguishable from the results in the original data.

\subsection{Outlier reweighting}

\begin{figure*}[t]
	\centering
	\includegraphics[width=0.9\textwidth]{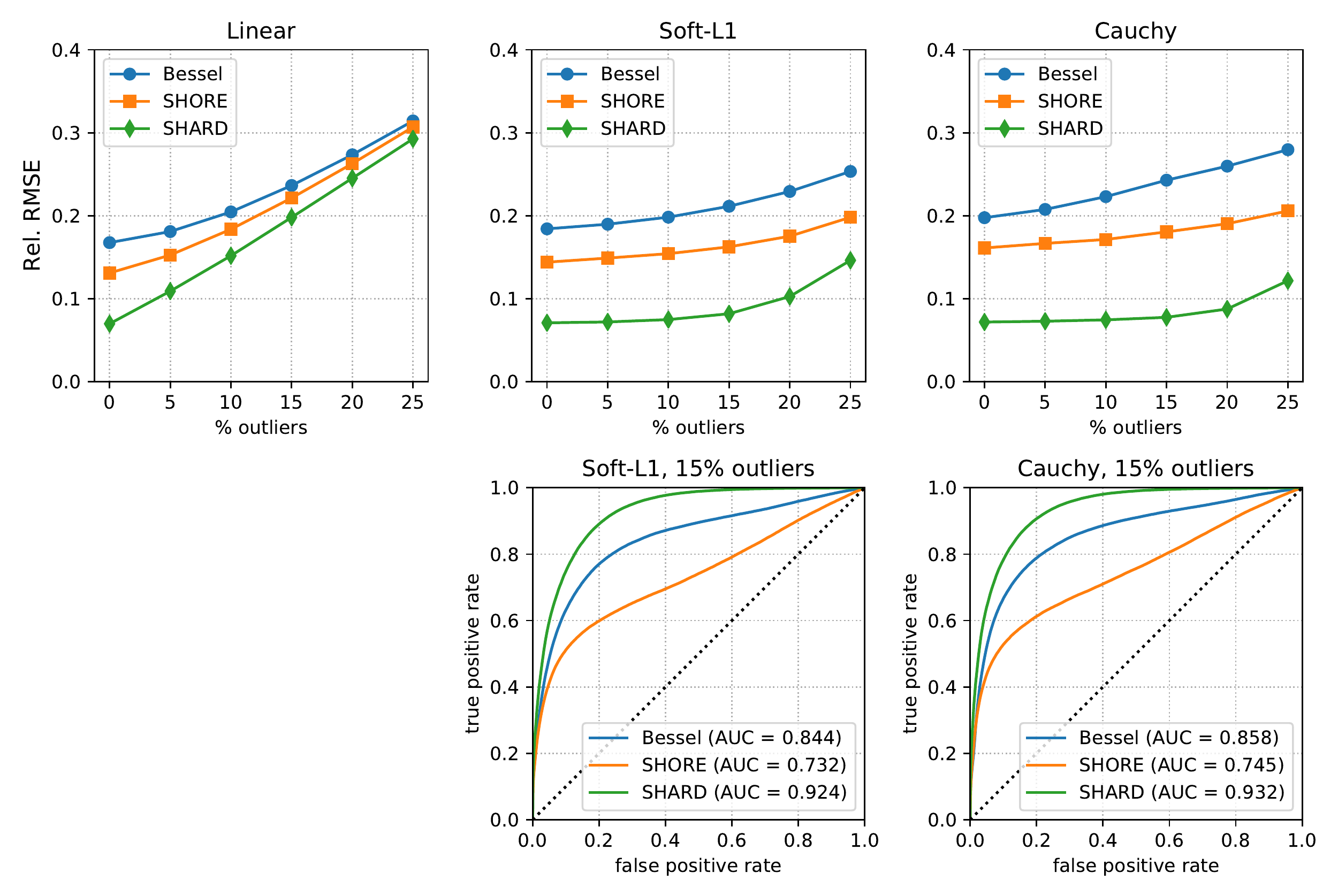}
	\caption{Performance of robust fitting under different simulated outlier probabilities in dataset~1 using basis rank $r=50$. Graphs in the top row show the relative RMSE between the original data and the predictions after linear least squares fitting, and after iterative reweighted least squares with Soft-$L_1$ and Cauchy loss functions. The bottom row plots receiver operator characteristic (ROC) curves of the sample outlier weights in the robust fitting at a simulated outlier level of 15\%.}
	\label{fig:or}
\end{figure*}

The outlier robust fitting scheme using iteratively re\-weighted least squares with Soft-$L_1$ and Cauchy loss functions is evaluated and compared to linear least squares in data with simulated outliers. The results are shown in Fig.~\ref{fig:or} for samples drawn uniformly from dataset~1, and in Suppl.\,Fig.~\ref{fig:or2} for samples drawn from dataset~2. As expected, the RMSE of the linear least squares fit increases with increasing levels of simulated outliers. In accordance with the results of Fig.~\ref{fig:rmse}, SHARD outperforms the Bessel and SHORE bases at low outlier levels. At outlier levels $\geq 20\%$, the margin narrows and all representations are equally corrupted. The robust estimator with Soft-$L_1$ loss substantially improves the RMSE at outlier levels over 5\%, at the expense of a small performance penalty in uncorrupted data. This improvement is stronger in SHARD than in the alternative representations. The benefit of a Cauchy loss function is comparatively smaller in the Bessel and SHORE representations and slightly larger in SHARD, widening the performance margin between them. In all experiments, SHARD achieved higher accuracy in outlier robust fitting than the other representations.

In addition, we validated the sample weights, assigned during iteratively reweighted least squares, in relation to the simulated outlier mask using receiver operator characteristic (ROC) curves shown in the bottom row of Figs.~\ref{fig:or} and \ref{fig:or2}. ROC curves plot sensitivity versus specificity, parametrized by the weight threshold value. The area under the ROC curve, which provides an aggregate measure of classifier performance, is highest with the SHARD basis in all experiments, indicating improved capability to discriminate outlier samples.

%%%%%%%%%%%%%%%%%%%%%%%%%%%%%%%%%%%%%%%%%%%%%%%%%%%%%%%

\section{Discussion}

The SHARD basis provides a linear, orthonormal representation of $q$-space dMRI data, with potential interest for a wide range of techniques and applications that exploit redundancy or rank-reduction. This decomposition builds on minimal prior assumptions, namely antipodal symmetry of the $q$-space signal and isotropic (rotation-invariant) frequency spectra of the radial basis. In contrast to the SHORE and Bessel function bases, this data representation is calibrated to the data at hand, akin to other data-driven and blind source separation techniques \cite{Christiaens2017}. Our results have shown that this data-driven basis construction reduces the residuals in fitting multi-shell data, especially in the low-rank regime. SHARD hence provides a more accurate data representation for matched number of parameters.

As shown in Section~\ref{sec:theory}, the resulting basis spans the complete microstructural information encoded in the signal and can thus provide a lossless data representation. Its primary advantage, however, is for compact, lossy data representation. This property was verified experimentally by the two-fold RMSE reduction at low rank in Fig.~\ref{fig:rmse}, and further illustrated for unsupervised multi-tissue decomposition in Fig.~\ref{fig:odfs}, showing that the low-rank representation facilitates highly efficient data compression. SHARD hence provides a suitable and efficient characterization of the measured dMRI signal and the underlying microstructure.

A closer examination of Fig.~\ref{fig:basis} reveals first of all that the covariance matrices qualitatively resemble those of simulated white matter in Fig.~\ref{fig:simulation}b, though not of rank\nobreakdash-1 and with an additional imprint of noise on the diagonal. Secondly, the derived basis functions, whilst obtained without any pre-imposed model, have desirable physical properties also found in the SHORE basis: the $n=1$ components of $\ell=0,2,4,\ldots$ capture continuously higher $b$-values and the $n > 1$ components behave as orthogonal harmonics of increasing frequency. The SH regularization in equation~\eqref{eq:shfit}, although not technically needed, can help reducing adverse effects from physiological noise in the $b \leq \unitfrac[400]{s}{mm^2}$ shells (Suppl.\,Fig.~\ref{fig:shpower}) that are otherwise captured in the basis functions. Its weight $\gamma$ thus offers control over unphysical (non-diffusion) effects in the data-driven basis construction that can be explored in relation to each application.

The main limitation of the current formalism is the requirement for spherically sampled data. This limitation stems from the fact that SHARD decomposes the radial $q$-space domain independently after projecting each shell to the SH basis. Spline interpolation or other numerical techniques can nevertheless be used to predict the signal between shells, either by interpolating the basis vectors in \eqref{eq:interp} or by interpolating the rows of matrix \eqref{eq:matstruct1} before decomposition. Future work can investigate means of extending the signal representation to arbitrary $q$-space sampling. This would require extending the current radial decomposition to the joint radial and angular domain, inherently encompassing the spherical harmonics projection. 
In addition, a joint groupwise data representation currently requires all subjects to be acquired with the same $b$-value sampling scheme, due to the discrete nature of the basis functions. However, when this requirement is fulfilled, a common group-level basis could be directly derived by extending matrix \eqref{eq:matstruct1} with voxel data of multiple subjects. This too can be subject of future work.

The SHARD representation has numerous potential applications for dMRI processing and analysis. This work has already illustrated a first, straightforward example in outlier re\-weighting using robust regression. Outlier robust fitting exploits optimal rank-reduction of the SVD, leveraging oversampling in $q$-space to detect outliers that lie too far from their predicted values. One can then either reject or down\-weight these outlier samples in further analysis, or proceed directly with the robustly-fitted SHARD signal representation. Another potential application is motion and distortion correction, in which data of individual volumes or slices are aligned with a common, iteratively-updated registration target, usually obtained as a low-rank prediction of the data. A proof-of-concept of using SHARD in motion correction was recently shown in \cite{Christiaens2018ismrm}. The results of the leave-one-out cross-validation in Fig.~\ref{fig:pred} indicate that it is indeed advantageous to predict individual slices and volumes from all data, rather than from a single-shell.

Furthermore, it is worth noting that the advantages of SHARD for dimensionality reduction can also be applied across the radial $q$-domain. This can be a useful property for selecting the number of shells and their most discriminating $b$-values in multi-shell protocol design, as was already successfully demonstrated on isotropic ($\ell=0$) averages \cite{Tournier2015}. SHARD can extend this approach to incorporate the higher harmonic orders.

Finally, it has not escaped our notice that the rank\nobreakdash-1 decomposition of Eq.~\eqref{eq:voxevd} suggests a generalised set of rotation-invariant signal features at the voxel or patch level. Indeed, these local \emph{SHARD features} may offer a potential means for predicting microstructure parameters, akin to recent methods that rely on signal features extracted per shell \cite{Novikov2018a, Reisert2017}, whilst exploiting the covariance in the combined multi-shell data and thus potentially less prone to noise. A patch-level SHARD decomposition may also be combined with random matrix theory for determining the optimal rank threshold in image denoising \cite{Veraart2016}. Future work can explore the merits of these extended applications.

%%%%%%%%%%%%%%%%%%%%%%%%%%%%%%%%%%%%%%%%%%%%%%%%%%%%%%%

\section{Conclusion}

SHARD provides a complete, orthogonal representation of the dMRI signal, tailored to the spherical geometry of $q$-space and calibrated to the data at hand. Rank-reduced SHARD decomposition outperformed model-based alternatives tested, whilst maximally capturing the micro- and meso-structural information in the signal. As such, SHARD is better suited for applications that require low-rank data predictions, such as outlier rejection and motion correction.

%%%%%%%%%%%%%%%%%%%%%%%%%%%%%%%%%%%%%%%%%%%%%%%%%%%%%%%

%\section*{Acknowledgements}

%%%%%%%%%%%%%%%%%%%%%%%%%%%%%%%%%%%%%%%%%%%%%%%%%%%%%%%

\bibliographystyle{IEEEtran}
\bibliography{references}

%%%%%%%%%%%%%%%%%%%%%%%%%%%%%%%%%%%%%%%%%%%%%%%%%%%%%%%
% SUPPLEMENTARY MATERIAL

\clearpage
\renewcommand\thefigure{S.\arabic{figure}}    
\setcounter{figure}{0}
\renewcommand\thetable{S.\arabic{table}}    
\setcounter{table}{0}

\begin{figure*}
	\centering
	\includegraphics[width=.9\textwidth]{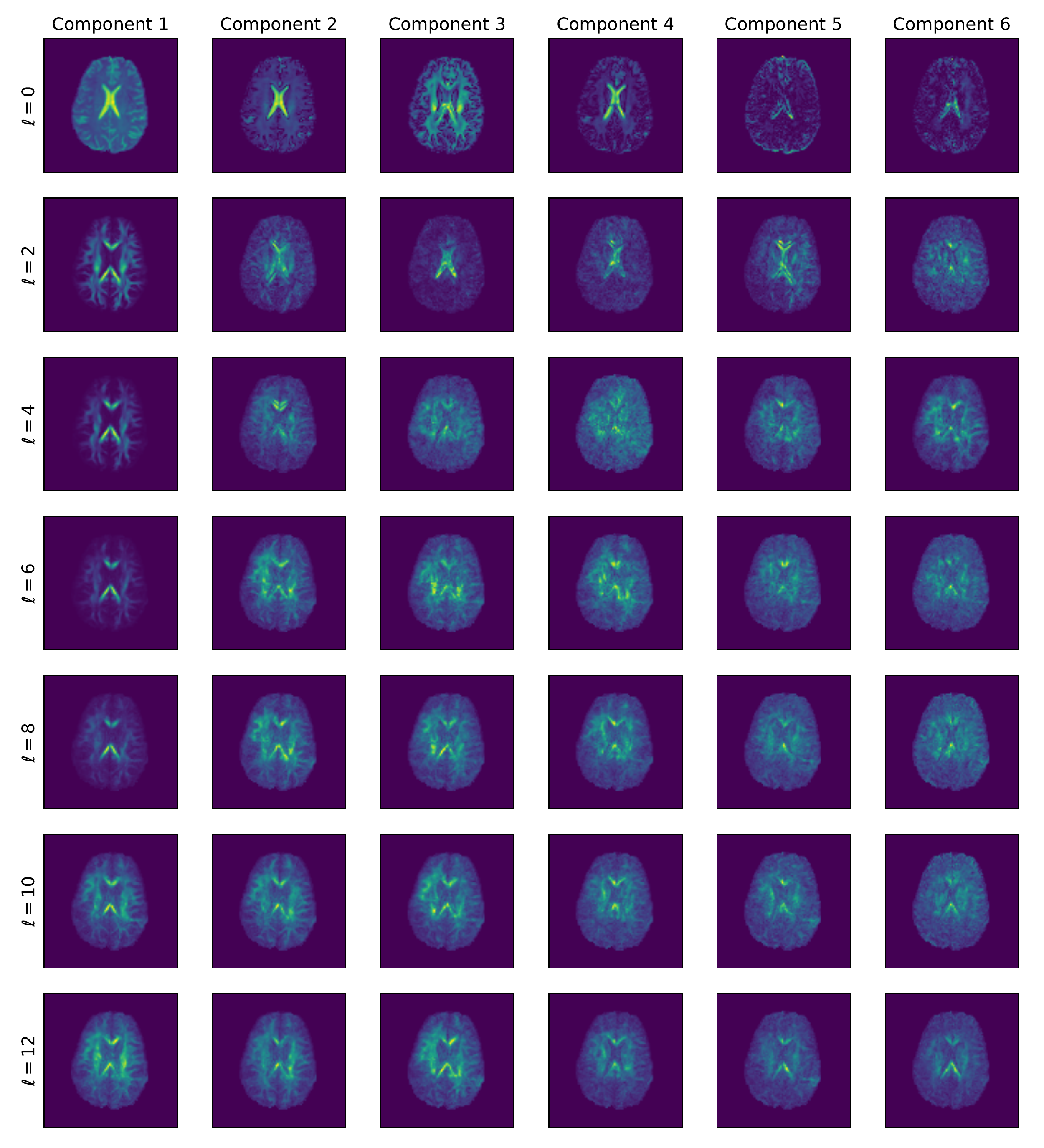}
	\caption{SHARD components of the multi-shell data for each harmonic band $\ell = 0, 2, \ldots, 12$ in dataset~1. The spatial maps depict the RMS power across the SH phase $m$ in each voxel. Only the 6 leading components are shown.}
	\label{fig:svdcomponents}
\end{figure*}

\begin{figure*}
	\centering
		\includegraphics[width=.9\textwidth]{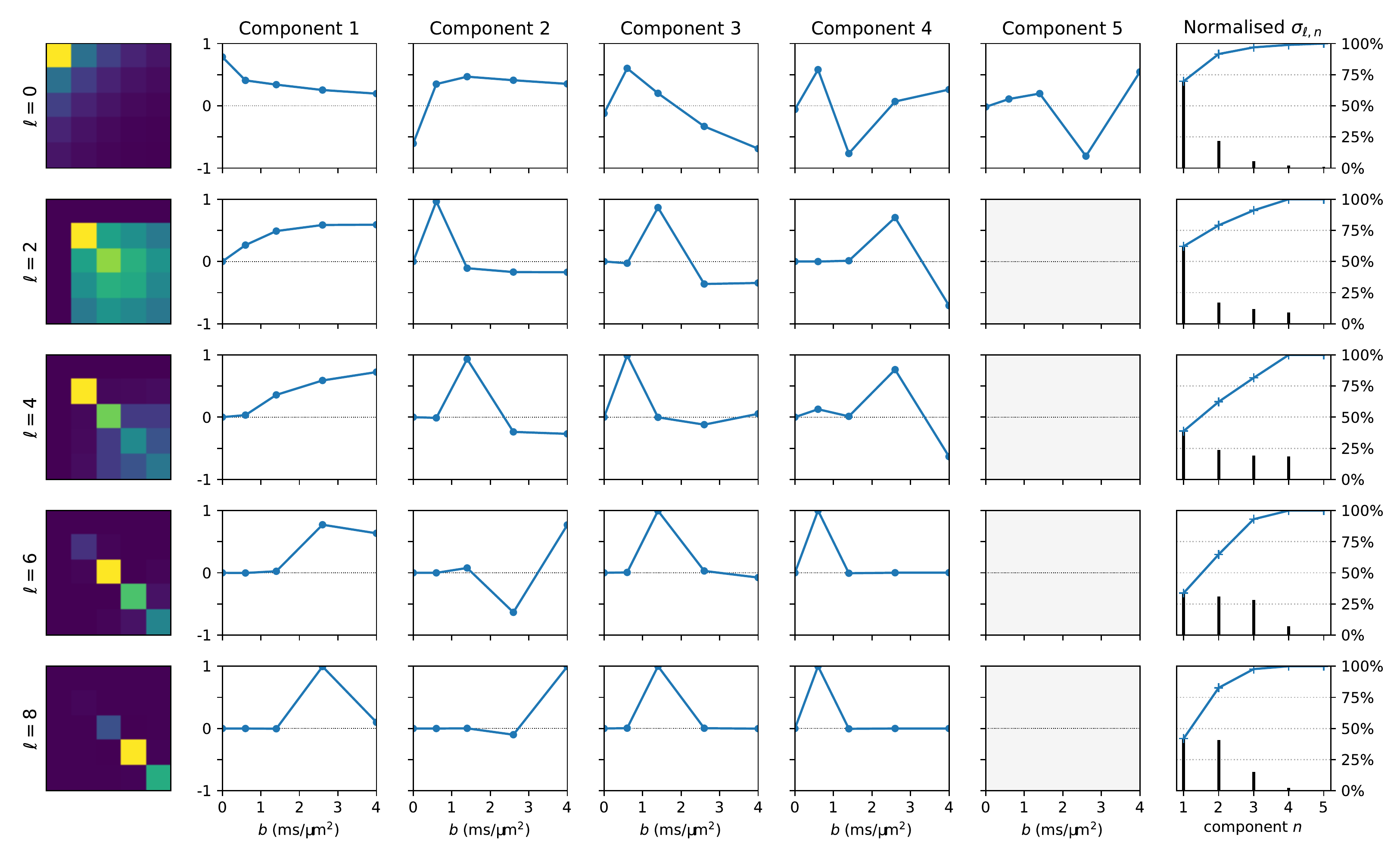}
	\caption{SHARD decomposition of dataset~2. Rows correspond to spherical harmonic bands of order $\ell = 0, 2, \ldots, 8$ The left column illustrates square, symmetric matrices $\tilde{\mat{S}}_\ell \, \tilde{\mat{S}}_\ell^\top$, showing the covariance between shells in each harmonic band. The middle columns plot the decomposition basis vectors, corresponding to the eigenvectors of the matrices on the left. Shaded boxes indicate components with $\sigma_{\ell,n}=0$, defined only by their orthogonality to the leading (data-driven) components and excluded from basis construction. The plots in the right column depict the singular values of each component in decreasing order and their cumulative sum.}
	\label{fig:basis2}
\end{figure*}

\begin{figure*}
	\centering
	\includegraphics[width=.8\textwidth]{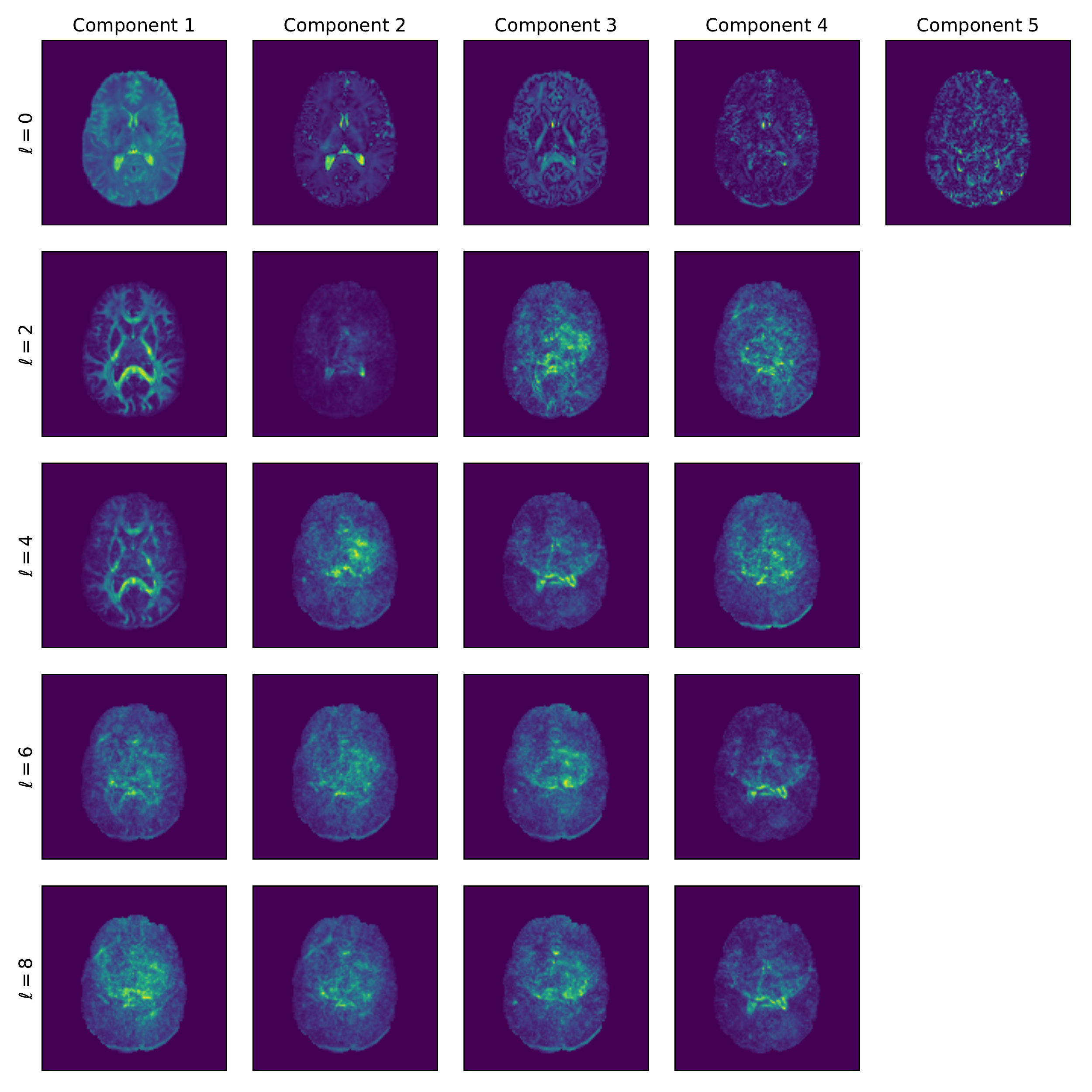}
	\caption{SHARD components of the multi-shell data for each harmonic band $\ell = 0, 2, \ldots, 8$ in dataset~2. The spatial maps depict the RMS power across the SH phase $m$ in each voxel.}
	\label{fig:svdcomponents2}
\end{figure*}

\begin{table}
\caption{Effect sizes $\varepsilon_j$ of all components $R_\ell^n(q)$ in sorted order, and the cumulative basis rank for dataset~1. The asterisk in the last column indicates the optimal rank as selected with leave-one-out cross-validation.}\label{tbl:effect1}
\centering
\begin{tabular}{c c r r c}
\toprule
$\ell$ & $n$ & effect $\sqrt{\varepsilon_j}$ & rank $r$ \\
\midrule
0 & 0 & 229.308 & 1   \\
0 & 1 &  48.075 & 2   \\
0 & 2 &  17.846 & 3   \\
0 & 3 &  11.081 & 4   \\
2 & 0 &   7.336 & 9   \\
0 & 4 &   4.303 & 10  \\
0 & 5 &   2.212 & 11  \\
4 & 0 &   2.089 & 20  \\
0 & 6 &   1.597 & 21  \\
2 & 1 &   1.042 & 26  \\
0 & 7 &   0.772 & 27  \\
0 & 8 &   0.711 & 28  \\
2 & 2 &   0.695 & 33  \\
6 & 0 &   0.680 & 46 & $\ast$  \\
2 & 3 &   0.629 & 51  \\
2 & 4 &   0.593 & 56  \\
0 & 9 &   0.502 & 57  \\
2 & 5 &   0.496 & 62  \\
0 & 10 &  0.439 & 63  \\
2 & 6 &   0.403 & 68  \\
4 & 1 &   0.382 & 77  \\
4 & 2 &   0.371 & 86  \\
2 & 7 &   0.354 & 91  \\
2 & 8 &   0.344 & 96  \\
4 & 3 &   0.342 & 105 \\
4 & 4 &   0.333 & 114 \\
4 & 5 &   0.317 & 123 \\
8 & 0 &   0.308 & 140 \\
4 & 6 &   0.306 & 149 \\
6 & 1 &   0.283 & 162 \\
6 & 2 &   0.273 & 175 \\
6 & 3 &   0.264 & 188 \\
6 & 4 &   0.243 & 201 \\
8 & 1 &   0.237 & 218 \\
4 & 7 &   0.229 & 227 \\
8 & 2 &   0.209 & 244 \\
6 & 5 &   0.203 & 257 \\
\bottomrule
\end{tabular}
\end{table}

\begin{table}
\caption{Effect sizes $\varepsilon_j$ of all components $R_\ell^n(q)$ in sorted order, and the cumulative basis rank for dataset~2. The asterisk in the last column indicates the optimal rank as selected with leave-one-out cross-validation.}\label{tbl:effect2}
\centering
\begin{tabular}{c c r r c}
\toprule
$\ell$ & $n$ & effect $\sqrt{\varepsilon_j}$ & rank $r$ \\
\midrule
0 & 0 & 86.981 & 1 \\
0 & 1 & 27.253 & 2 \\
0 & 2 & 6.697 & 3 \\
2 & 0 & 4.783 & 8 \\
0 & 3 & 2.482 & 9 \\
4 & 0 & 1.333 & 18 \\
0 & 4 & 1.320 & 19 & $\ast$ \\
2 & 1 & 1.297 & 24 \\
2 & 2 & 0.922 & 29 \\
4 & 1 & 0.807 & 38 \\
2 & 3 & 0.688 & 43 \\
6 & 0 & 0.659 & 56 \\
4 & 2 & 0.659 & 65 \\
4 & 3 & 0.634 & 74 \\
6 & 1 & 0.606 & 87 \\
8 & 0 & 0.564 & 104 \\
\bottomrule
\end{tabular}
\end{table}

\begin{figure*}
	\centering
	\includegraphics[width=0.9\textwidth]{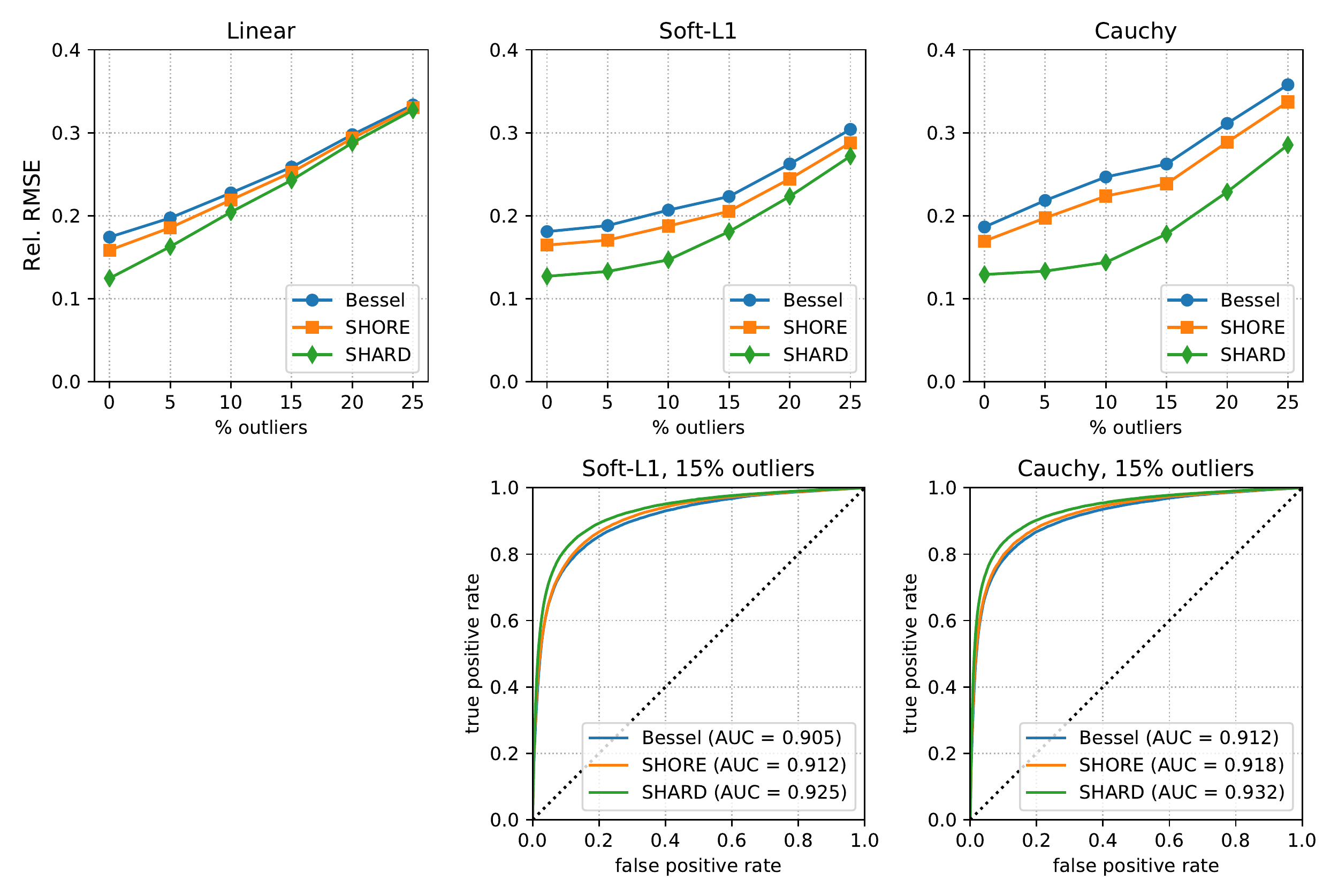}
	\caption{Performance of robust fitting under different simulated outlier probabilities in dataset~2 using basis rank $r=22$. Graphs in the top row show the relative RMSE between the original data and the predictions after linear least squares fitting, and after iterative reweighted least squares with Soft-$L_1$ and Cauchy loss functions. The bottom row plots receiver operator characteristic (ROC) curves of the sample outlier weights in the robust fitting at a simulated outlier level of 15\%.}
	\label{fig:or2}
\end{figure*}

\begin{figure*}
	\centering
	\includegraphics[width=0.8\textwidth]{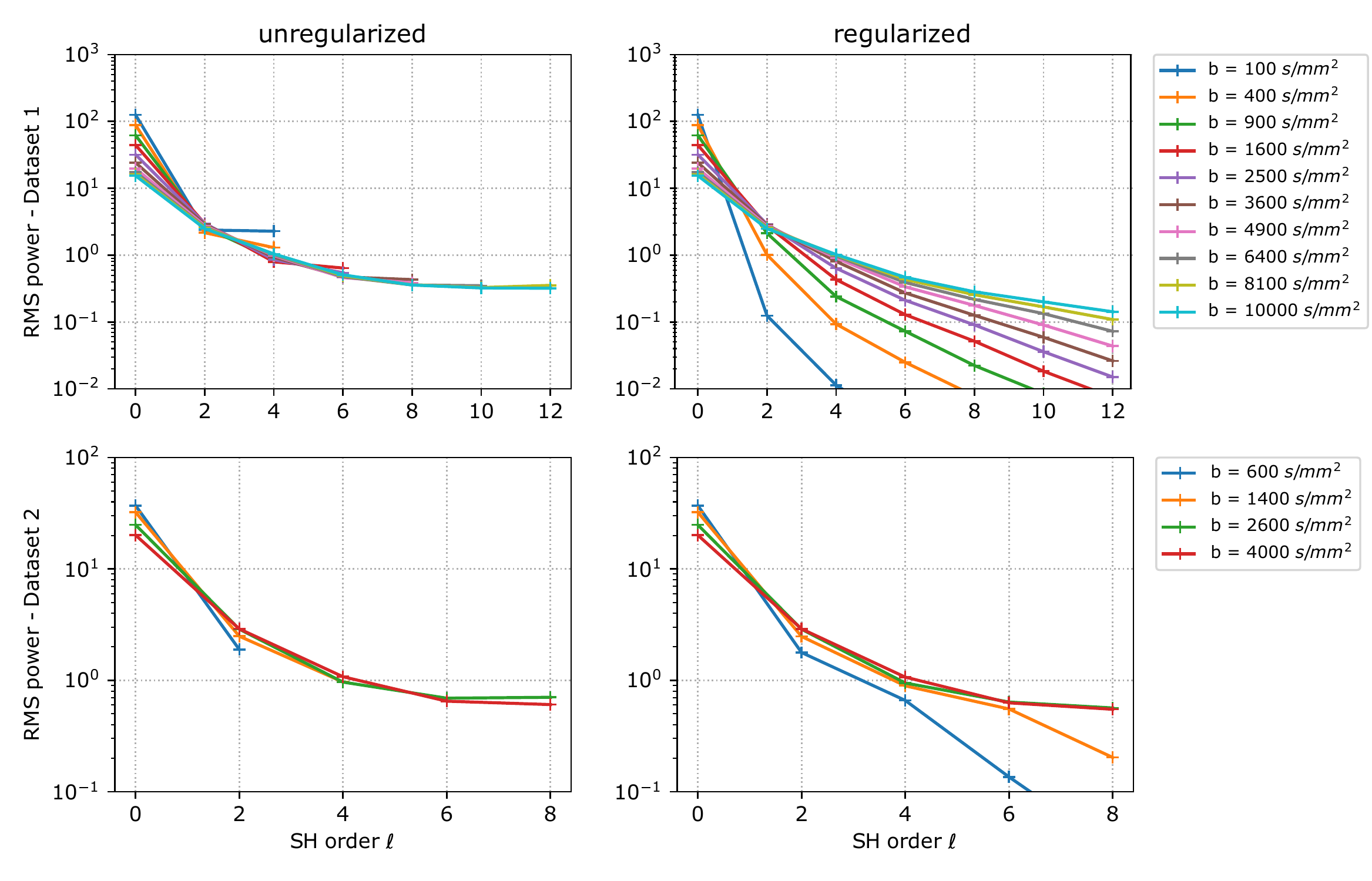}
	\caption{Power spectra in multi-shell spherical harmonics (SH) fitting without and with regularization. Conventional unregularized fitting (left column) requires truncating $\ell_\text{max}$ in each shell to the maximum overdetermined order, which can lead to the staircasing effect seen in dataset~1. Adding Laplace-Beltrami regularization (right column) enables fitting all shells at the same $\ell_\text{max}$. We can also observe that the regularizer effectively suppresses high-frequency components in the $b=\unitfrac[100]{s}{mm^2}$ and $b=\unitfrac[400]{s}{mm^2}$ shells of dataset~1, attributed to physiological noise rather than dMRI signal.}
	\label{fig:shpower}
\end{figure*}

\end{document}